# Realization of a hole-doped Mott insulator on a triangular silicon lattice

Fangfei Ming,[1] Steve Johnston,[1] Daniel Mulugeta,[1] Tyler S. Smith,[1] Paolo Vilmercati,[1,2] Geunseop Lee,[3] Thomas A. Maier,[4] Paul C. Snijders,[5,1] and Hanno H. Weitering[1,5]

[1]*Department of Physics and Astronomy, The University of Tennessee, Knoxville, Tennessee 37996*

[2]*Joint Institute of Advanced Materials at The University of Tennessee, Knoxville, TN 37996*

[3]*Department of Physics, Inha University, Inchon 402-751, Korea*

[4] *Computational Science and Engineering Division and Center for Nanophase Materials Sciences, Oak Ridge National Laboratory, Oak Ridge, Tennessee 37831*

[5]*Materials Science and Technology Division, Oak Ridge National Laboratory, Oak Ridge, Tennessee 37831*

**The physics of doped Mott insulators is at the heart of some of the most exotic physical phenomena in materials research including insulator-metal transitions, colossal magneto-resistance, and high-temperature superconductivity in layered perovskite compounds. Advances in this field would greatly benefit from the availability of new material systems with similar richness of physical phenomena, but with fewer chemical and structural complications in comparison to oxides. Using scanning tunneling microscopy and spectroscopy, we show that such a system can be realized on a silicon platform. Adsorption of one-third monolayer of Sn atoms on a Si(111) surface produces a triangular surface lattice with half-filled dangling bond orbitals. Modulation hole-doping of these dangling bonds unveils clear hallmarks of Mott physics, such as spectral weight transfer and the formation of quasi-particle states at the Fermi level, well-defined Fermi contour segments, and a sharp singularity in the density of states. These observations are remarkably similar to those made in complex oxide materials, including high-temperature superconductors, but highly extraordinary within the realm of conventional *sp*-bonded semiconductor materials. It suggests that exotic quantum matter phases can be realized and engineered on silicon-based materials platforms.**

It has long been surmised that semiconductor surfaces could be ideal platforms for studying correlated electron phenomena in low-dimensional electron systems [1-8]. In particular, the reduced coordination of the surface atoms of covalently bonded semiconductors implies the existence of a two-dimensional (2D) periodic array of partially filled dangling bonds. Most of these systems are susceptible to band Jahn-Teller or charge ordering instabilities, however, which can lead to a complete restructuring of the surface [9]. From a chemical perspective, these types of instabilities are driven by the system's tendency to saturate or eliminate partially filled dangling bonds through orbital rehybridization and/or the formation of new bonds. From the physics perspective, these 2D systems undergo spontaneous symmetry breaking transitions that are driven by strong electron-phonon coupling, wiping out most or all density of states (DOS) near the Fermi level $E_F$. In principle, the charge ordering instability competes with magnetic or superconducting instabilities, although this type of physics seems mostly limited to the realm of transition metal oxides with narrow *d*-bands and moderate-to-strong electron correlations [10]. The beauty of those systems is that the various interactions and resulting order parameters can be tuned via e.g. controlled doping experiments, providing access to important quantum matter phases, including unconventional superconductivity [10].

The 'α-phases' of Sn and Pb on Si(111) or Ge(111) are believed to exhibit similarly rich physics [3, 4, 6-8, 11, 12]. These phases are formed by adsorbing a 1/3 monolayer (ML) of Sn or Pb on the Si(111) or Ge(111) surface in an ordered ($\sqrt{3}\times\sqrt{3}$)*R*30° arrangement. In this geometry, the substrate's dangling bonds are passivated while each Sn or Pb adatom contributes one half-filled dangling bond per unit cell (see Fig. 1(a)). With the exception of Sn on Si(111), these systems all undergo a charge ordering transition at low temperature [3, 4, 8, 12]. A recent angle-resolved photoemission (ARPES) study indicated that Sn on Si(111) is an antiferromagnetic Mott insulator below 200 K [13], or perhaps a magnetic band or Slater insulator [14]. If the Mott scenario can be confirmed, Sn on Si(111) (henceforth $\sqrt{3}$-Sn), would represent one of the closest experimental realizations of the spin ½ triangular-lattice antiferromagnetic Heisenberg model. Such a system could host exotic spin states, such as a chiral spin state or a quantum spin liquid [15], and become superconducting with doping [7].

A critical first step toward accomplishing this goal would be the successful introduction of excess charges into the $\sqrt{3}$-Sn dangling bond lattice without introducing structural and/or chemical disorder. Here, we employ a modulation doping scheme by using Si(111) substrates with various bulk doping levels [16]. Because the charge reservoir in the dangling bond surface state is in thermodynamic equilibrium with that of the doped bulk material, the use of heavily-doped p-type (n-type) substrates should in principle produce a hole-doped (electron-doped) surface layer (see Note 1 in Ref. [17]). We used n-type silicon substrates with room temperature (RT) resistivities of 0.002, 0.03, and 3 Ω·cm, and p-type (B-doped) substrates with RT resistivities of 0.03, 0.008, 0.004 and 0.001 Ω·cm. They are labeled according to their doping type and resistivity, for instance "n-0.002" or "p-0.008" (smaller numbers represent higher bulk doping levels). The substrate with the largest boron content, p-0.001, is labeled B$\sqrt{3}$. Here, segregated boron atoms form a ($\sqrt{3}\times\sqrt{3}$)*R*30° superstructure below the topmost Si layer, which corresponds to the highest achievable doping level with our modulation doping approach [16].

The silicon substrates were annealed to temperatures over 1000 °C in ultrahigh vacuum (UHV) so as to prepare atomically clean Si(111)7x7 and Si(111)($\sqrt{3}\times\sqrt{3}$)*R*30°-B substrates for the growth of the $\sqrt{3}$-Sn structure. Sn atoms were deposited from a thermal effusion cell while keeping the substrate temperature at 550 °C, which resulted in the formation of a well-ordered $\sqrt{3}$-Sn structure. An Omicron low-temperature scanning tunneling microscope (STM) was used for atomic resolution imaging and spectroscopy of the (un)doped √3-Sn surfaces. Additional details can be found in the supplementary text [17].

Using the modulation doping scheme, we were able to hole-dope the √3-Sn structure up to about 10%. We were unable to create an electron-doped √3-Sn structure on the n-type substrates, however. This difficulty is related to the fact that high-temperature annealing in UHV leads to dopant conversion in the subsurface region of n-type Si [18-20]. To check this scenario for our samples, we performed current-voltage (I(V)) measurements on the n- type substrate with the STM. Our data confirm the existence of a p-type inversion layer below the surface of the n-3 sample (see Note 2 in Ref. [17]), indicating that its surface is hole doped. The subsurface layer of n-0.03 and n-0.002 sample is partially compensated but remains n-type.

Fig. 1(b) shows a STM topographic image of the √3-Sn surface. The atoms seen in this image are the Sn adatoms forming a (√3×√3)$R$30° superstructure relative to the underlying Si(111) substrate [4], (Fig. 1(a)). The image remains the same, regardless of the doping level or doping type, except for the fact that the √3-Sn domain size decreases with increased hole doping (see Note 3 in Ref. [17]). We find no evidence for structural distortions in these images.

Fig. 2(a) shows a series of $dI/dV$ tunneling spectra, recorded at 77 K, representing the local density of states (LDOS) of the √3-Sn surface for different doping levels [22]. The bottommost n-0.002 spectrum agrees well with the spectrum expected for an undoped Mott-Hubbard insulator [6,7] with a lower and an upper Hubbard band (LHB/UHB) centered at about -0.4 V and +0.5 V, respectively, and a 0.2 eV Mott gap in between. This assignment is fully consistent with previous experiments [6,13,23] and theoretical calculations [11,13,24]. The Mott insulating nature of the n-0.002 sample indicates that the excess electrons from the partially compensated substrate are trapped by localized defect states (see Note 2 in Ref. [17]).

By switching to the n-0.03 substrate, the system enters the hole-doped regime. This hole doping is signaled by the formation of a shoulder slightly above $E_F$, which evolves into a quasi-particle peak (QPP) whose tail crosses the Fermi level as more holes are added to the surface. Note that the QPP is already visible for the n-3 substrate, consistent with the established presence of the p-type inversion layer below the surface. It is centered at about + 0.1 eV and its intensity is greatest for the B√3-Sn surface. The origin of this peak must be electronic in nature. It cannot be associated with a structural distortion as the STM images of the doped and undoped systems are identical. While hole doping clearly metallizes the √3-Sn surface, the zero bias conductance in STM remains suppressed for all doping levels, except for the B√3-Sn case (see Note 4 in Ref. [17]). This is reminiscent of the pseudo-gap feature observed on lightly doped cuprate and iridate compounds [25, 26]. However, unlike the cuprates and iridates, the spectral features of the √3-Sn system are remarkably uniform spatially (see Note 5 in Ref. [17]). In contrast, doped cuprate and iridate materials usually exhibit competing nanophases such as striped or checkerboard-like charge ordered regions [27]. This suggests that these competing orders are material-specific and are not intrinsic to the single-band Hubbard model.

It is evident from Fig. 2(a) that the growth of the QPP with hole-doping is accompanied by a reduction of spectral weight of the Hubbard bands. The spectral weights can be estimated by integrating the areas under the UHB, LHB and QPP for each doping level, keeping in mind that the total spectral weight should be independent of doping. This procedure is described in detail in Supplementary Note 6 in Ref. [17] and the result is shown in Fig. 2(b). The spectral weight of the QPP reaches a maximum of about 20% on the B√3-Sn sample, which corresponds to a hole doping level of about 10% [17]. The observed spectral weight transfer is a very distinctive hallmark of Mott physics [28, 29], and is well reproduced by our Dynamical Cluster Approximation (DCA) calculations [30], shown in Fig. 2(c), where we used the hopping parameters and Hubbard U from Ref. [13]. Importantly, the observed spectral weight transfer rules out the recently proposed Slater insulator scenario for the undoped system [14], where the band gap arises from magnetic ordering and associated band folding. Doping a Slater band insulator would result in a repositioning of the chemical potential but does not lead to a redistribution of spectral weight [29].

At high hole doping levels, quasi-particles contributing to the spectral weight near $E_F$ should acquire momentum dispersion and develop a well-defined Fermi surface contour. The Fermi contour can be acquired by measuring the quasi-particle interference (QPI) patterns in $dI/dV$ STM imaging near zero bias [31]. These experiments are necessarily limited to low temperature and, consequently, to only the most conductive samples. Fig. 3(a) shows a topographic STM image of the B√3-Sn system, acquired with a tunneling bias of - 5 mV at 4.4 K. The Sn adatoms are perfectly equivalent in both filled and empty state imaging, again ruling out the presence of a charge-density modulation [3] or structural rippling at 4.4 K. Fig. 3(b) is the simultaneously acquired $dI/dV$ image. The latter reveals a significant spatial modulation of the differential tunneling conductance $g(\boldsymbol{r},V) = dI(\boldsymbol{r},V)/dV$ as quasi-particles scatter at defects and, consequently, between different **k**-points on the constant energy contour $k(E=eV)$. Its Fourier transform, $g(\boldsymbol{q},V)$, shown in Fig 3(d), represents the power spectrum of the density modulation in Fig. 3(b). Note that for zero bias, the scattering vector $\boldsymbol{q}$ = 2$k_F$ ± $\boldsymbol{G}$, where $\boldsymbol{G}$ is a reciprocal lattice vector of the (√3×√3)*R*30° superstructure (see Note 7 in Ref. [17]), meaning that the Fermi surface can be reconstructed from the QPI Fourier transform [31].

Filled-state $g(\boldsymbol{q},V)$ maps are shown in Figs. 3(c) and 3(d); empty state maps are shown in Figs. 3(e) and 3(f). The most intense spots are the Bragg peaks of the (√3×√3)*R*30° surface structure, which allows us to determine the magnitude and direction of the scattering vectors. It should be noted that the clarity of the scattering features in these maps is extraordinary, which is true even for the raw data (see Fig. S8 in Ref. [17]). The differential conductance maps furthermore reveal the presence of approximately equilateral triangles pointing toward the origin. They are highlighted in green. The outer edges of these triangles form a warped hexagonal constant energy contour, highlighted in red. The other edges are scattering replicas involving a reciprocal lattice vector translation. The triangles gradually shrink going from negative to positive bias, meaning that the corresponding scattering vectors disperse with energy. The dispersion relation $\boldsymbol{q}(E)$ is plotted along the $\overline{\Gamma K}$ and $\overline{\Gamma M}$ directions of the surface Brillouin zone in Figs. 3(k) and 3(l), respectively.

We simulated the $g(\boldsymbol{q}, V)$ maps using a single-band model obtained from density functional theory (DFT) calculations [13] and the T-Matrix formalism, assuming a single point-like potential scatterer (see [32] and Note 8 in Ref. [17]). The resulting simulated maps are shown in Figs. 3(g-j), alongside the corresponding experimental maps. Both the triangular features (green) and warped contours (red) are well reproduced. The experimental and theoretical dispersion relations in Figs. 3(k) and 3(l) are also very similar. The warping of the triangular features in the simulated power spectrum can be traced to the warped hexagonal shape of the Fermi contour, as seen in DFT calculations [13, 4]. The experimental triangles have flatter sides, however, which is particularly evident in panels 3(f) and 3(j): the constant energy contour at +10 meV appears perfectly nested. This flattening of the constant energy contours as compared to the DFT results reflects the renormalization of the quasi-particle dispersion due to the Hubbard interaction. Most importantly, the existence of a Fermi surface contour in the hole-doped system that is reminiscent of the non-interacting bands of the √3-Sn structure, is conclusive evidence that the origin of the QPP should be attributed to the spectral weight transfer of a hole-doped Mott insulator. Clearly, the QPP feature is intrinsic to the √3-Sn structure and cannot be related to the formation of some impurity band, which should have a completely different dispersion, if any.

Closer inspection of the QPP offers a preview of what might happen if the hole-doping level could be pushed beyond the ten percent level achieved for the B√3-Sn surface. The inset of Fig. 2(d) shows the recognizable QPP just above zero bias in $dI/dV$ spectrum; this time the spectrum is recorded at 4.4 K. Note the very sharp spike on top of the QPP. The zoomed-in spectra in Fig. 2(d) clearly reveal the emergence of a singularity in the DOS ~6 meV below the Fermi level, which is becoming increasingly prominent as the temperature decreases. This feature can be attributed to a van Hove singularity in the DOS [33], associated with the saddle point in the quasi-particle band dispersion near the $\overline{\text{M}}$-point [13]. Indeed, the experimental and theoretical dispersions in Fig. 3(l) exhibit a kink near the M-point at about -5 meV, consistent with this saddle point. Further hole doping is expected to drive the singularity up to the Fermi level, which almost certainly would give rise to an electronic instability. In fact, such an instability may already exist below the current measurement temperature minimum of 4.4 K.

In light of the strong on-site Coulomb repulsion and approximate nesting conditions, one could envision two types of instability. The first would be similar to that envisioned for e.g. heavily doped graphene [34]. With its hexagonal Fermi contour and saddle points at the $\overline{\text{M}}$-points, graphene would be susceptible to a *d*-wave superconducting instability if only the van Hove singularity at ~2 eV [35] from the Dirac point could be driven to the vicinity of Fermi level, as it is here. Likewise, both doped Mott physics and nesting have been considered essential for *d*-wave pairing in high temperature superconductors [36, 37]. The alternative scenario would be a competing charge-order or spin-density wave instability. Charge ordering in related systems nucleates at defects sites [38] for which there is no evidence here. On the other hand, the near perfect nesting conditions with three 120° rotated nesting vectors could give rise to a non-coplanar chiral spin order [39]. While the system would not exhibit long-range magnetic order, chirality would exist at finite temperatures, producing a spontaneous quantum Hall effect [39]. More doping studies will be needed to determine if these scenarios can indeed be realized on the silicon platform.

**Acknowledgement:** This work was primarily funded by the National Science Foundation under grant DMR 1410265. S.J. is partially funded by the University of Tennessee's Science Alliance Joint Directed Research and Development (JDRD) program, a collaboration with Oak Ridge National Laboratory. T.M. was sponsored by the Laboratory Directed Research and Development Program of Oak Ridge National Laboratory, managed by UT-Battelle, LLC, for the U. S. Department of Energy. G.L. acknowledges supports from the National Research Foundation of Korea (NRF) funded by the Korean government (MSIP) (NRF-2017R1A2B2003928).

**References**

1. J. E. Northrup, J. Ihm, and M. L. Cohen, Phys. Rev. Lett. **47**, 1910 (1981).
2. J. Ortega, F. Flores, and A. Levy Yeyati, Phys. Rev. B **58**, 4584 (1998).
3. J. M. Carpinelli, H. H. Weitering, E. W. Plummer, and R. Stumpf, Nature **381**, 398 (1996).
4. J. M. Carpinelli, H. H. Weitering, M. Bartkowiak, R. Stumpf, and E. W. Plummer, Phys. Rev. Lett. **79**, 2859 (1997).
5. H. H. Weitering, X. Shi, P. D. Johnson, J. Chen, N. J. DiNardo, and K. Kempa, Phys. Rev. Lett. **78**, 1331 (1997).
6. S. Modesti, L. Petaccia, G. Ceballos, I. Vobornik, G. Panaccione, G. Rossi, L. Ottaviano, R. Larciprete, S. Lizzit, and A. Goldoni, Phy. Rev. Lett. **98**, 126401 (2007).
7. G. Profeta, and E. Tosatti, Phys. Rev. Lett. **98**, 086401 (2007).
8. R. Cortés, A. Tejeda, J. Lobo-Checa, C. Didiot, B. Kierren, D. Malterre, J. Merino, F. Flores, E. G. Michel, and A. Mascaraque, Phys. Rev. B **88**, 125113 (2013).
9. C. B. Duke, Chem. Rev. **96**, 1237-1260 (1996).
10. J. Orenstein, and A. J. Millis, Science **288**, 468-474 (2000).
11. P. Hansmann, T. Ayral, L. Vaugier, P. Werner, and S. Biermann, Phys. Rev. Lett. **110**, 166401 (2013).
12. J. Avila, A. Mascaraque, E. G. Michel, M. C. Asensio, G. LeLay, J. Ortega, R. Pérez, and F. Flores, Phys. Rev. Lett. **82**, 442 (1999).
13. G. Li, P. Höpfner, J. Schäfer, C. Blumenstein, S. Meyer, A. Bostwick, E. Rotenberg, R. Claessen, and W. Hanke, Nat. Commun. **4**, 1620 (2013).

14. J. H. Lee, X. Y. Ren, Y. Jia, and J. H. Cho, Phys. Rev. B **90**, 125439 (2014).
15. P. Fazekas, and P. W. Anderson, Philos. Mag. **30**, 423 (1974).
16. F. Ming, D. Mulugeta, W. Tu, T. S. Smith, P. Vilmercati, G. Lee, Y. T. Huang, R. D. Diehl, P. C. Snijders, and H. H. Weitering, Nat. Commun. **8**, 14721 (2017).
17. See Supplemental Material at [URL will be inserted by publisher] for [band alignment measurement; spectral weight transfer estimation; sample growth and surface morphology; temperature dependent Mott transitions; spatial homogeneity; processing procedure of the QPI data; details about theoretical calculations].
18. M. Liehr, M. Renier, R. A. Wachnik, and G. S. Scilla, J. Appl. Phys. **61**, 4619-4625 (1987).
19. S. Bensalah, J-P. Lacharme, and C. A. Sébenne, Phys. Rev. B **43** 14441 (1991).
20. J. D. Mottram, A. Thanailakis, and D. C. Northrop, J. Phys. D **8**, 1316 (1975).
21. M. Rashidi, M. Taucer, I. Ozfidan, E. Lloyd, M. Koleini, H. Labidi, J. L. Pitters, J. Maciejko, and R. A. Wolkow, Phys. Rev. Lett. **117**, 276805 (2016).
22. J. Tersoff, & D. R. Hamann, Phys. Rev. Lett. **50**, 1998 (1983).
23. A. Charrier, R. Pérez, F. Thibaudau, J. M. Debever, J. Ortega, F. Flores, and J. M. Themlin, Phys. Rev. B **64**, 115407 (2001).
24. S. Schuwalow, D. Grieger, and F. Lechermann. Phys. Rev. B **82**, 035116 (2010).
25. P. Cai, W. Ruan, Y. Peng, C. Ye, X. Li, Z. Hao, X. Zhou, D.-H. Lee, and Y. Wang, Nat. Phys. **12**, 1047 (2016).
26. I. Battisti, K. M. Bastiaans, V. Fedoseev, A. De La Torre, N. Iliopoulos, A. Tamai, E. C. Hunter, R. S. Perry, J. Zaanen, F. Baumberger and M. P. Allan, Nat. Phys. **13**, 21 (2017).
27. P. A. Lee, N. Nagaosa, and X. G. Wen, Rev. Mod. Phys. **78**, 17 (2006).
28. E. Louis, F. Flores, F. Guinea, and J. Tejedor, J. Phys. C **16**, L39 (1983).
29. H. Eskes, , M. B. J. Meinders, G. A. Sawatzky, Phys. Rev. Lett. **67**, 1035 (1991).
30. T. Maier, M. Jarrell, T. Pruschke, M. H. Hettler, Rev. Mod. Phys. **77**, 1027 (2005).
31. L. Petersen, P. Hofmann, E. W. Plummer, and F. Besenbacher, J. Electron. Spectrosc. Relat. Phenom. **109**, 97-115 (2000).
32. H. Alloul, J. Bobroff, M. Gabay, and P. J. Hirschfeld, *Rev. Mod. Phys.* **81**, 45 (2009).
33. A. Piriou, N. Jenkins, C. Berthod, I. Maggio-Aprile, and Ø. Fischer, Nat. Commun. **2**, 221 (2011).
34. R. Nandkishore, L. S. Levitov, and A. V. Chubukov, Nat. Phys. **8**, 158-163 (2012).
35. R. Saito, G. Dresselhaus, and M. S. Dresselhaus, *Physical Properties of Carbon Nanotubes* (Imperial College Press, London, 1998).
36. R. S. Markiewicz, J. Phys. Chem. Solids **58**, 1179-1310. (1997).
37. D. M. Newns, C. C. Tsuei, and P. C. Pattnaik, Phys. Rev. B **52**, 13611 (1995).
38. H. H. Weitering, J. M. Carpinelli, A. V. Melechko, J. Zhang, M. Bartkowiak, and E. W. Plummer, Science **285**, 2107-2110 (1999).
39. I. Martin, and C. D. Batista, Phys. Rev. Lett. **101**, 156402. (2008).

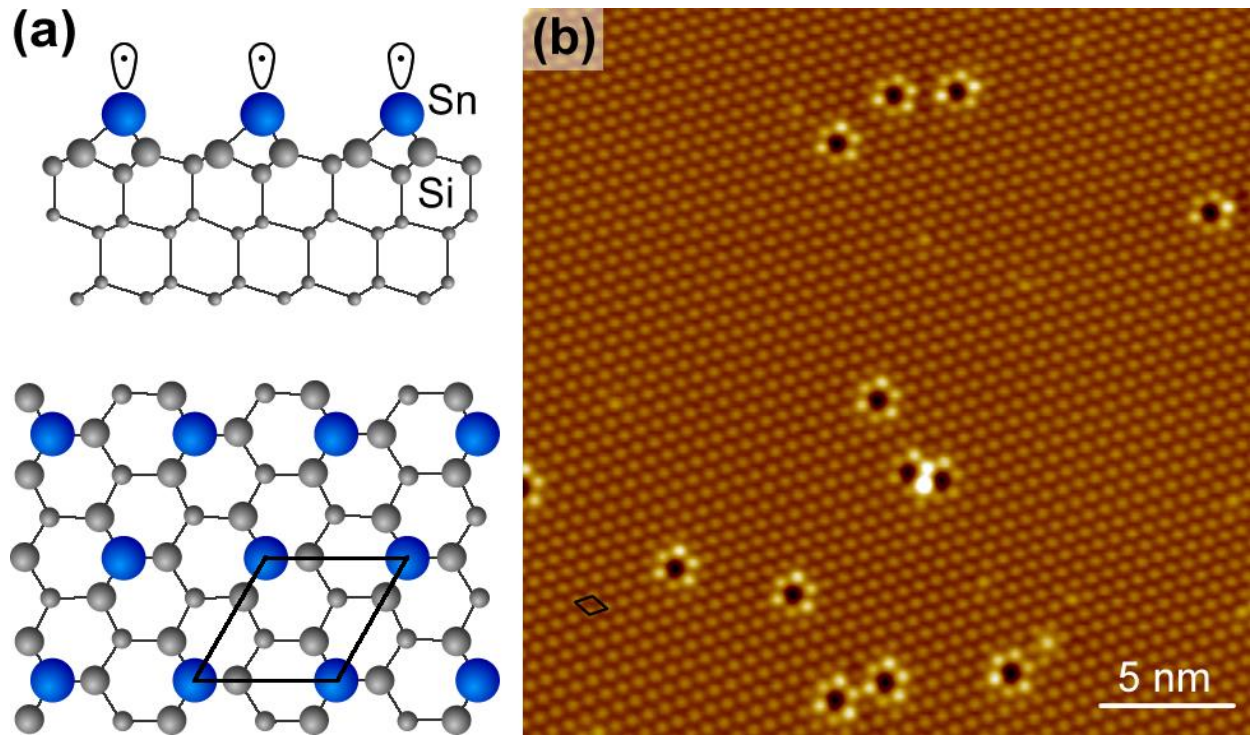

FIG. 1 Structure and STM topography of the √3-Sn Mott insulating phase of Sn on Si(111). (a) Side view (top) and top view (bottom) of the √3-Sn structure. The Sn adatoms are colored in blue, each contributing a nominally half-filled dangling bond orbital. There are no other dangling bonds present in this structure. (b) STM image of the √3-Sn structure on a n-0.002 Si(111) substrate, recorded at 77 K ($V_s$ = -2 V, $I_t$ = 0.1 nA). The individual Sn atoms are clearly resolved. Dark features surrounded by hexagonal rings of bright adatoms likely correspond to substitutional Si defects in the Sn layer. Diamonds in panels (a) and (b) mark the (√3×√3)$R$30° unit cell.

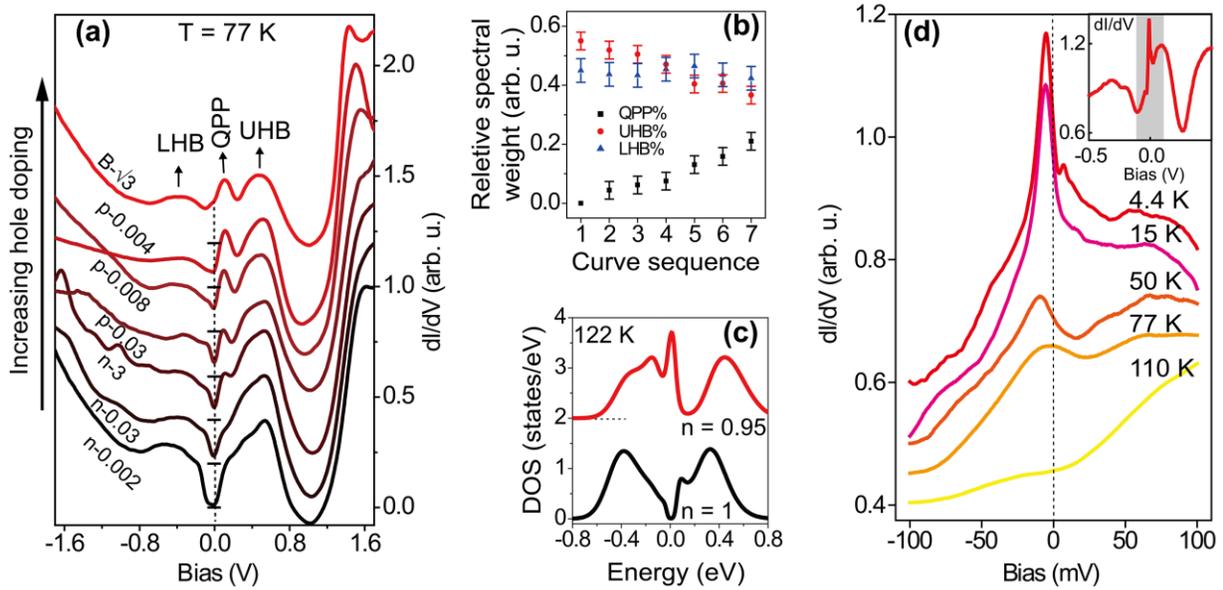

FIG. 2 (a), $dI/dV$ spectra ($\propto$ LDOS) measured at 77 K for the √3-Sn surface grown on various n-type and p-type Si substrates. The spectra are normalized to a bulk feature at 1.6 V. Horizontal tick marks on the dashed vertical line at zero bias indicate the origin of the differential conductance axis for each curve. Note the negative differential resistance between 0.8 eV and 1.2 eV, consistent with other tunneling spectroscopy studies of localized dangling bond states [21]. (b) Relative intensities of the UHB, LHB and QPP according to the spectral integration procedure in Supplementary Note 6 [17]. (c) Density of states obtained from DCA calculations for the undoped and five-percent hole-doped √3-Sn surfaces at 122 K. (d), Temperature dependent $dI/dV$ spectra of the B√3-Sn sample, showing an emerging van Hove singularity at about 6 mV below the Fermi level. The inset in panel (d) shows the 4.4 K spectrum on a larger bias scale. On this scale, the van Hove singularity is visible as a sharp spike riding on top of the QPP. The bias range displayed in the main panel is marked in gray. The spectra in (a), (c) and (d) are offset vertically for clarity.

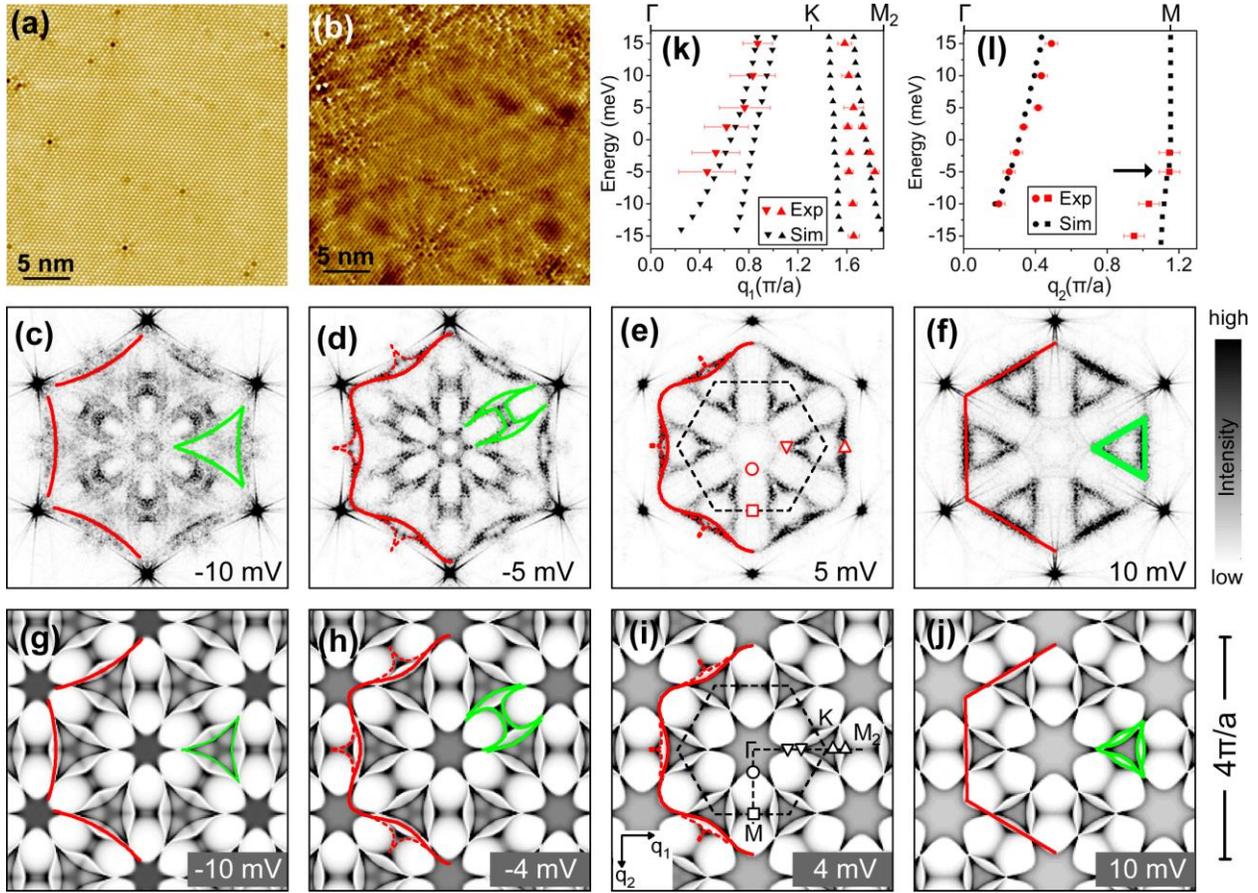

FIG. 3 Quasi particle interference patterns of the hole doped √3-Sn Mott insulator on the B√3 substrate. (a) STM topographic image and (b) simultaneously acquired differential conductance map $g(\mathbf{r},V)$ at -5 mV and 4.4 K. (c-f), Fourier transforms $g(\mathbf{q},V)$ of the conductance maps $g(\mathbf{r},V)$ acquired at different voltages. Note that panel (d) is the Fourier transform of image (b). The $g(\mathbf{q},V)$ images have been rotated such that a $\overline{\Gamma K}$ direction runs horizontally while a $\overline{\Gamma M}$ direction runs vertically, as indicated in panels (e) and (i). The $g(\mathbf{q},V)$ images were symmetrized, employing the three-fold rotational lattice symmetry in conjunction with the mirror symmetry in k-space (see Fig. S8 in Ref. [17]). The six outer spots are the Bragg spots of the (√3×√3)$R$30° real space lattice. (g-j) Simulated $g(\mathbf{q},V)$ using the non-interacting band structure from Ref. [13] and applying the standard T-Matrix formalism for a single scattering defect [23]. Note the warped hexagonal contour in the experimental and theoretical power spectra, indicated in red. This is the $|\mathbf{q}|=2|\mathbf{K}(E)|$ contour, where $\mathbf{K}(E)$ is a constant energy contour in the surface state band structure. (k&l) Dispersion relation E($\mathbf{q}$) extracted from the bias dependent $g(\mathbf{q},V)$ images along the $\overline{\Gamma KM}$ direction (k) and $\overline{\Gamma M}$ direction (l). Red symbols represent the experimental data points (see Fig. S9 in Ref. [17]). Black dots are the theoretical results. The kink (labelled with an arrow in (l)) in the band dispersion near the M-point at -6 meV is consistent with the presence of the van Hove singularity in Fig. 2d.

# Supplemental Material for

# Realization of a hole-doped Mott insulator on a triangular silicon lattice

Fangfei Ming,[1] Steve Johnston,[1] Daniel Mulugeta,[1] Tyler S. Smith,[1] Paolo Vilmercati,[1,2] Geunseop Lee,[3] Thomas A. Maier,[4] Paul C. Snijders,[5,1] and Hanno H. Weitering[1,5]

[1]*Department of Physics and Astronomy, The University of Tennessee, Knoxville, Tennessee 37996*

[2]*Joint Institute of Advanced Materials at The University of Tennessee, Knoxville, TN 37996*

[3]*Department of Physics, Inha University, Inchon 402-751, Korea*

[4] *Computational Science and Engineering Division and Center for Nanophase Materials Sciences, Oak Ridge National Laboratory, Oak Ridge, Tennessee 37831*

[5]*Materials Science and Technology Division, Oak Ridge National Laboratory, Oak Ridge, Tennessee 37831*

## Note 1: Band alignment

To determine the direction of the charge transfer between the surface and the bulk, as well as the nature of the space charge layer beneath the surface, one must know the location of the Fermi level at the surface and inside the bulk, relative to the position of the bulk band edges. The Fermi level position in the bulk depends on the doping level and doping type, as well as temperature, and can be calculated using simple textbook formula [1]. The Fermi level location at the surface is unknown but can be determined from the binding energy of the Si 2p core level as measured with x-ray photoelectron spectroscopy (XPS). The procedure is as follows.

Si 2p core level spectra were recorded *in-situ* with XPS, using Al K$\alpha$ radiation ($\hbar\omega$ = 1486.6 eV). The spectra were measured at room temperature on both the n-0.002 and p-0.03 substrates, before and after the growth of a uniform √3-Sn reconstruction. The UHV flash anneal of these silicon substrates produced a well-ordered Si(111)-(7×7) reconstruction. The distance between the Fermi level $E_F$ and valence band maximum (VBM) *at the 7x7 surface* is $E_F - E_{VBM}$ = 0.63 ± 0.05 eV [2]. The measured Si 2*p* core level binding energies are 99.1 ± 0.05 eV and 98.9 ± 0.05 eV for the n-0.002n and p-0.03 Si(111)-(7x7) surfaces, respectively. For the √3-Sn reconstructions grown on the n-0.002 and p-0.03 substrates, the Si 2*p* core levels are shifted 0.12 ± 0.05 eV towards lower binding energy, due to the realignment of the Fermi level at the surface. These data, together with the calculated Fermi level positions deep inside the bulk [1], enable us to determine the band bending diagram for the √3-Sn reconstructions on the two Si substrates and, consequently, the direction of the charge transfer between the surface and the bulk.

Fig. S1(a) summarizes our findings. For the √3-Sn grown on n-0.002 and p-0.03 substrates, $E_F - E_{VBM}$ = 0.51 ± 0.07 eV and 0.52 ± 0.07 eV, respectively. Note that within the margin of error, the Fermi level is pinned at the surface (that is, $E_F - E_{VBM}$ is the same for both n and p type substrates). Using these band-bending data, one could in principle estimate the excess charge transferred to the surface by calculating the concentration of ionized dopants in the space charge layer below the surface within the abrupt depletion layer approximation and assuming a homogeneous doping distribution [3]. Using the nominal doping level of the bulk wafer and the band bending parameters obtained from XPS, we obtain a surface doping level below 1% for all substrates used. This estimate appears to contradict the prominent spectral weight transfer observed in the case of p-type wafers. The B dopants in the p-type wafers tend to segregate to the surface, however, significantly increasing the dopant concentration near the surface up to a maximum concentration of 1/3 of a monolayer for the B-√3 reconstructed substrate. In this latter case, complete ionization and trapping of all holes at the surface would produce a hole doping level of one hole per Sn adatom. This doping level would, in turn, produce a band-insulating surface similar to the Si(111)(√3×√3)R30°-B surface. The metallic nature of the √3-Sn surface grown on the B√3 substrate indicates that the ionization of the B dopants near the surface is not complete and/or that the B concentration beneath the √3-Sn domains is less than 1/3 ML.

As pointed out in the main text, electron doping experiments were unsuccessful, which may be due to dopant redistribution in the subsurface region, as will be discussed in Note 2. Hence, it should be understood that the band diagrams in Fig. S1 are valid only if the dopant distribution in the bulk is homogeneous and if there are no charge compensating defects. Nonetheless, the location of the band edges at the surface relative to the Fermi level must be correct. This also allows us to indicate the bulk band edges in the tunneling spectra.

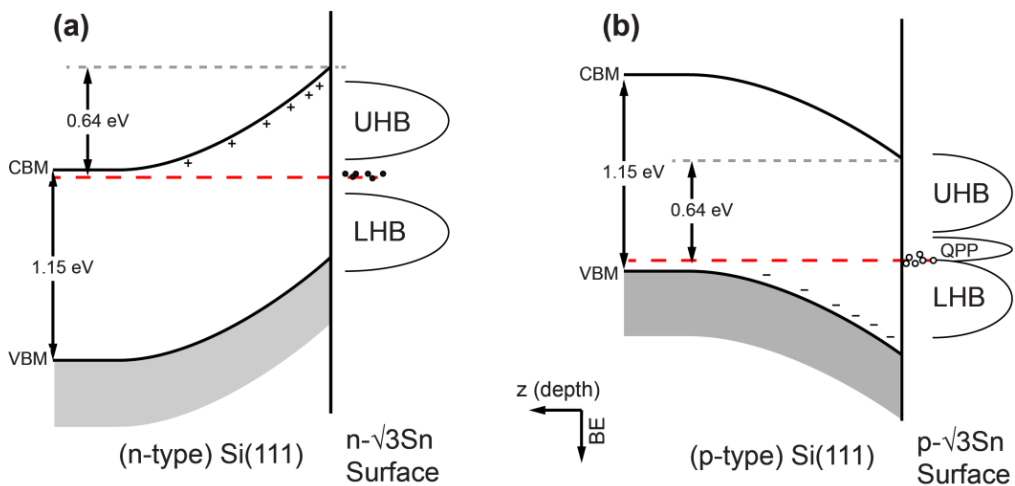

Figure S1 - Band alignment diagram for the √3-Sn surface grown on n-0.002 (a) p-0.03 (b) substrates, as determined from XPS measurements. Energies are relative to the Fermi energy (red dashed line). The surface state spectrum, i.e., the Hubbard bands and quasi particle peak, are also indicated schematically.

## Note 2: Hole doping mechanism for n-type substrates.

As discussed in conjunction with Fig. 2(b), the doping level of the √3-Sn surface appears to be shifted toward the hole doped regime. The surface of the most heavily doped n-type substrate (n-0.002) appears to be 'undoped' while the √3-Sn surface on the n-3 substrate is clearly hole doped. This observation can be understood as follows.

1. While boron atoms tend to segregate toward the surface upon high temperature annealing, the n-type dopants (e.g., As, Sb) diffuse deeper into the bulk. As shown in Refs. [4-6], annealing at ~ 1150 °C for 5 minutes reduces the subsurface concentration of As or Sb dopants by about one order of magnitude. Our samples are annealed at 1200 °C in order to remove the native silicon oxide and create a clean and well-ordered silicon substrate for the growth of the √3-Sn layer. We therefore expect that the n-type doping concentration near the surface will be significantly less than the nominal concentration of those wafers. In addition, UHV annealing leads to unavoidable boron contamination. According to Ref. [4], 5 minutes of annealing in UHV at 1150 °C produces a near-surface B concentration of approximately $8\times10^{17}$ cm$^{-3}$, which is similar to the boron content of moderately doped p-type Si. This situation could conceivably create an inversion layer near the surface if the near-surface n-type doping concentration would be less than that of the B contaminants. Indeed, from point-contact STS measurements, we find that the lightly doped n-type substrate (n-3: phosphorous doped with a nominal doping concentration of $1.5\times10^{15}$ cm$^{-3}$) has become p-type below the surface. Fig. S2 shows I(V) data from the n-0.03 and n-3 Si(111)-(7x7) surfaces. The I(V) curves of both samples clearly reveal rectifying behavior, due to the Schottky barrier formation between the metal tip and semiconductor sample. The Schottky barrier of the n-0.03 sample is forward-biased when the sample bias is negative, which means that the semiconductor side of the Schottky junction is n-type (see e.g. Ref. 1). Likewise, the n-3 sample is forward-biased at positive sample bias, meaning that the semiconductor side is p-type. This observation confirms the dopant compensation scenario discussed above.

2. The above scenario implies that all samples are p-type with the exception of the n-0.03 and n-0.002 samples. All p-type samples reveal a clear QPP in the STS spectra, characteristic of the hole-doped Mott insulator. The fact that the √3-Sn surface on the n-0.002 substrate appears to be undoped strongly suggests that the excess electrons from this partially compensated n-type substrate do not transfer to the Hubbard bands. A likely scenario is that these electrons are trapped by localized defects at the surface, which are usually considered responsible for the well-known phenomenon of Fermi level pinning [1, 3]. While the precise nature of those pinning defects require further study beyond the scope of the present paper, a likely candidate for electron trapping defect is the Si substitutional defect, i.e. a Si atom replacing a Sn atom.

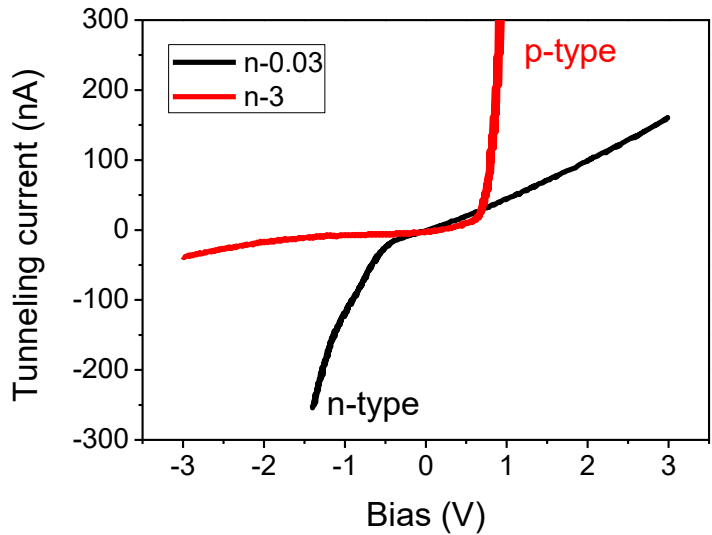

Figure S2 - I(V) curves measured on the Si(111)-(7x7) surface formed on the n-0.03 and n-3 substrate, measured at 77 K. The opposite rectification behavior of the red (n-3) and black (n-0.03) I(V) curves indicate that the Si near the surface is p-type (n-type) for the n-3 (n-0.03) substrates.

Several of those atomic defects are visible in Fig. 1 of the main text and their density is typically of the order of 2%, more than enough to trap all of the charge carriers originating from the space charge layer below the surface. (In Note 1, we estimated that those n-type charges would correspond to a doping level of less than 1 %).

Indeed, dI/dV images from individual defect often produce ring-like features when measured at 5 K, as shown in Fig. S3. These types of rings have been studied in great detail and are caused by electric-field induced defect ionization under the STM tip [7-9]. These features are not observable at 77 K, as the scanning field is well screened by the sample at these higher temperatures. The double ring feature in Fig. S3(c), obtained with negative tip bias, indicates that this substitutional Si defect is doubly occupied in the unperturbed state [7]. The presence of the defects thus provides a mechanism for 'self doping'. This naturally explains why the √3-Sn structure of the n-0.002 wafer is undoped. As a consequence, the √3-Sn on the n-0.03 sample must be weakly hole doped. Indeed, while the QPP has not yet fully developed, the Mott gap of the n-0.03 sample is beginning to fill. The upper limit for the hole density associated with self-doping must be of the order of one percent. Higher doping levels are only achievable through modulation hole doping.

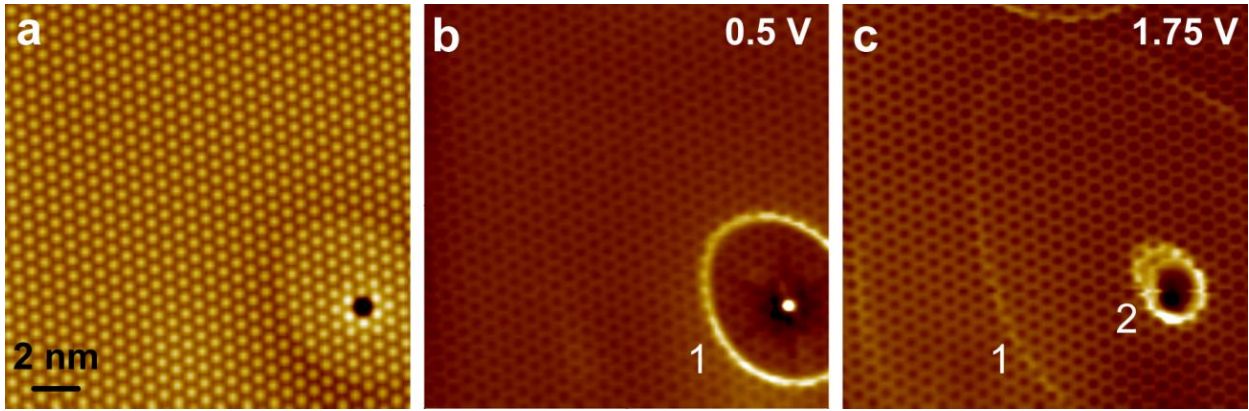

Figure S3 - Charge state transitions of a Si substitutional defect on the √3-Sn surface of the n-0.002 substrate, induced by the STM tip at 5 K. (a) STM image of a single surface defect. $V_s$ = 1.75V, and $I_t$ = 0.6 nA. (b) The simultaneously obtained dI/dV image showing a bright ring with the defect at its center. The appearance of such rings can be attributed to field induced ionization by the STM tip. It also affects the brightness in the topographical image near the defect(a). (c) dI/dV image scanned at the same location with $V_s$ = 1.75V, and $I_t$ = 0.6 nA. the defect is now at the center of two large rings. The first ring corresponds to the same transition observed in (b), while the second one corresponds to a removal of a second electron from the defect. The bright features at the upper or left edge is due to other surface defects out of the scanning area.

## Note 3: Growth of the (√3×√3)R30°-Sn surface reconstruction

Silicon single crystal wafers were prepared by flashing them up to 1200 °C in UHV, followed by a slow cool down to room temperature. This procedure produces atomically-clean well-ordered Si(111)-(7x7) substrates. To enrich the subsurface with boron dopants, the Si(111)(√3×√3)R30°-B surface was prepared by prolonged annealing of the p-0.001 substrate at ~ 1100 °C.

Deposition of 1/3 ML of Sn on the n-0.002, n-0.03, n-3, and p-0.03 Si(111)-(7x7) substrates at ~500 °C with several minutes of post-annealing at the same temperature produces a Si(111)(√3×√3)R30°-Sn surface, or √3-Sn surface for short. For Sn coverage slightly below 1/3 ML, the √3-Sn domains are somewhat disordered. Small excess amounts of Sn, on the other hand, produce large well-ordered √3-Sn domains with small coexisting Si(111)(2√3×2√3)R30°-Sn (or 2√3-Sn) domains. The latter appear to nucleate at the step edges, as shown in Fig. S4(a), and have a local coverage of about 1.2 ML [10]. To suppress defect formation in the √3-Sn phase, we always overexposed the surface. For accurate band bending measurements with XPS, which has very low spatial resolution, we made sure that the presence of 2√3-Sn domains was minimal.

The surface morphology is different for Sn on the heavily boron-doped p-0.004 and B-√3 substrates. Here, a substantial amount of boron has segregated to the surface prior to the deposition of Sn and some of the √3-Sn areas are highly defective. These defective √3-Sn (d-√3-Sn) areas have a clear boundary with the well-ordered √3-Sn domains, as seen in Fig. S4(b)&(c). STM imaging indicates that for these highly defective domains, Si atoms replace up to 80% of the Sn adatoms. In order to obtain well-ordered √3-Sn domains on the heavily B-doped substrates, it is necessary to increase the Sn coverage well beyond 1/3 ML. This inevitably results in the large scale appearance of the high-density 2√3-Sn phase. The optimal dosage for creating well-ordered √3-Sn domains depends on the doping level in the bulk substrate. In the most optimal conditions, we acquired a mixed phase surface where the competing 2√3-Sn phase covers about 25% or 85% of the p-0.004 and B-√3 substrate surfaces, respectively. On the B-√3 substrate, the majority 2√3-Sn phase fully encloses the smaller √3-Sn domains as seen in Fig. S4(c). The √3-Sn domains are generally anisotropic in shape and can grow up to about 100 nm in size when measured along their smallest dimension. We also observe some disordered islands on the Sn-covered p-0.004 and B-√3 substrates, as indicated with the "AI" labels in Fig. S4(b)&(c). Their density is highest on the B-√3 substrate.

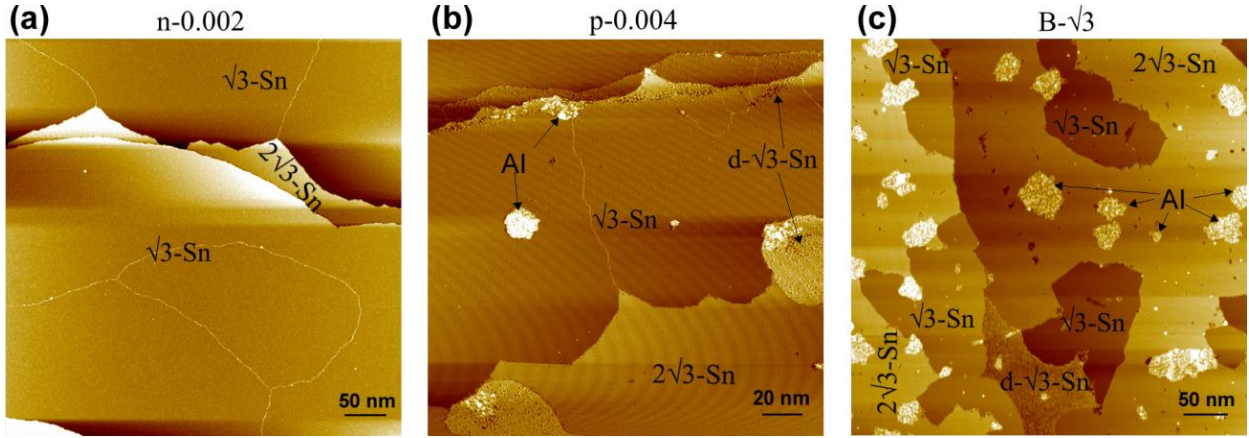

Figure S4 - Optimized surface morphologies for different doping levels in the bulk. STM images recorded at 77 K following the growth of Sn on the n-0.002 Si(111) substrate (a), p-0.004 substrate (b), and B√3 substrate (c). The estimated Sn coverages are 0.4 ML (a), 0.6 ML (b), and 0.9 ML (c). The various surface domains are labeled. Bright curved lines in panels (a) and (b) mark the boundaries between different √3-Sn domains. Amorphous islands are labeled with "AI", and "d-√3-Sn" stands for defective √3-Sn domains.

## Note 4: The Mott transition

As a consistency check with earlier work [11], we reproduced the temperature dependence of the Mott gap in Fig. S5. In agreement with Modesti *et al.* [11], we find that the zero-bias conductance of the n-0.002 sample vanishes below 100 K. The tunneling conductance still exhibits a soft gap or 'pseudo-gap' up to about 220 K, suggesting that 100 K should be regarded as the lower limit of the Mott transition temperature.

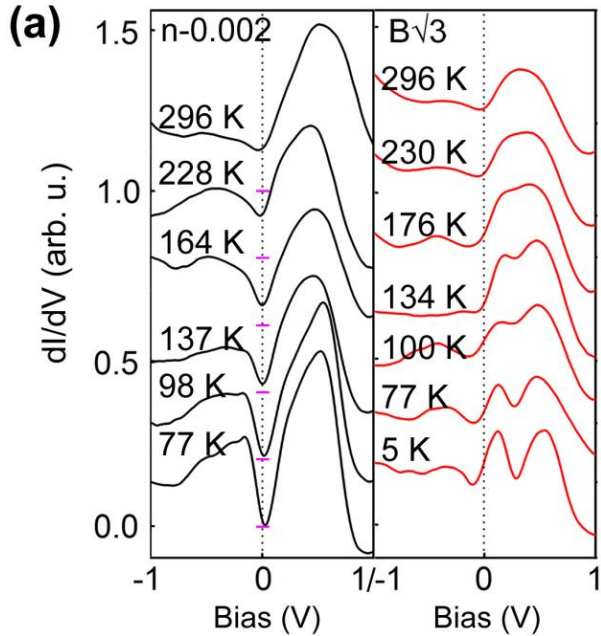

Figure S5 - Evolution of spectral features as a function of temperature. (a) Temperature dependent $dI/dV$ spectra for the √3-Sn structure on n-0.002 and on B√3 substrates, showing a gradual smearing of the low-energy spectral features. As in Fig. 2(a), all the curves are normalized to the peak at ~1.6V (beyond the plotted bias range). The curves are shifted vertically for clarity. The zeros in the conductance axis for the n-0.002 curves are indicated by the horizontal tick marks. The Mott gap of the n-0.002 sample gradually disappears at higher temperature. Importantly, the spectral features of the hole-doped system grown on the B√3 substrate, in particular the QPP, are also washed out above 200 K. This reinforces the correlation between the presence of the QPP and the opening of the Mott gap below 200 K. (b) Temperature dependent zero bias conductance for the √3-Sn surface on an n-0.002 substrate. Evidently, the metal-insulator transition sets in at about 100 K and is completed at about 200 K, consistent with Ref. 11.

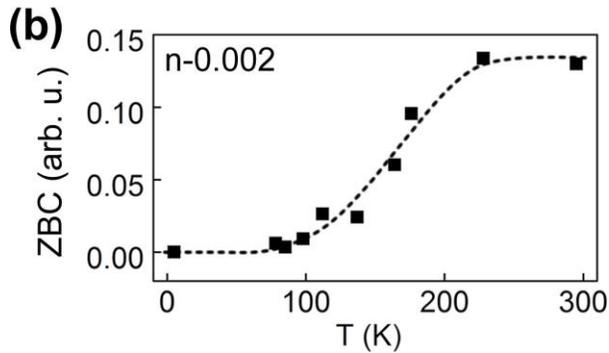

# Note 5: Spatial homogeneity of the low energy spectral features

Generally, STM/STS spectra measured on correlated oxides such as the high-$T_c$ cuprates display a large degree of inhomogeneity, which arises from short-range charge ordering and/or defects and dopants in the system [12]. Remarkably, we find that the √3-Sn surface is very homogeneous, both within the UHB and LHB, as well as within the QPP. We attribute this to the low density of defects in the Sn layer and the modulation doping scheme used here, which physically locates the dopant atoms away from the electronic states forming the low-energy sector.

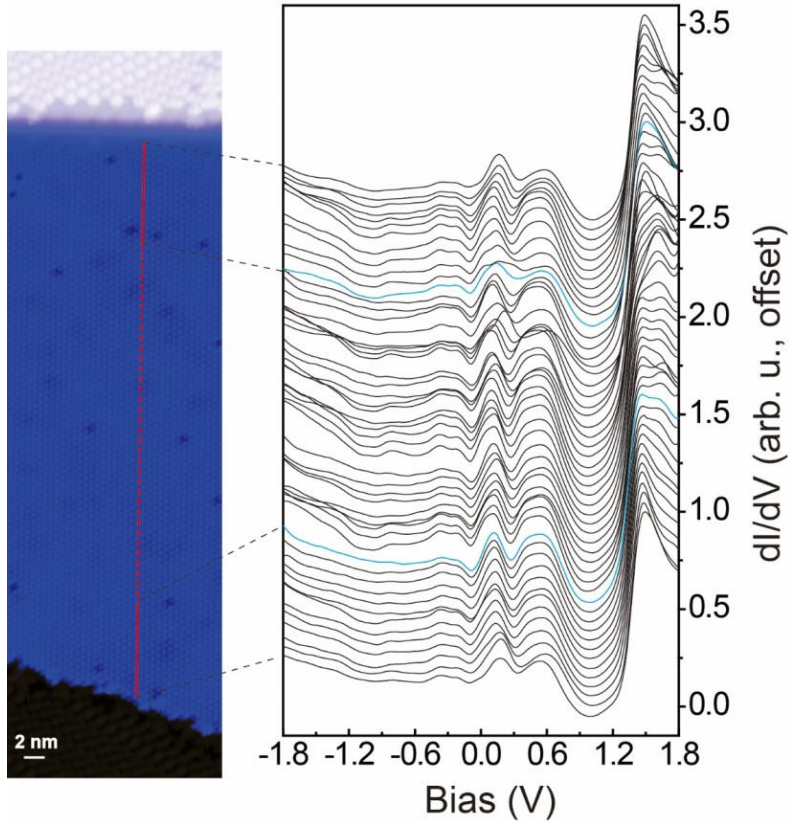

Figure S6 - Spectral homogeneity of the √3-Sn surface on the B√3 substrate. $dI/dV$ spectra (right) along a line on a wide terrace of the √3-Sn phase grown on a B√3 substrate (left). The √3-Sn terrace is bound by two step edges. The spectra, recorded at 5 K, show that the intensity and location of the low energy spectral features (±0.7 V, UHB, LHB, QPP) are constant across the surface. Along the solid line segments, one spectrum was recorded per unit-cell, whereas in the dashed line segment one spectrum was recorded every other unit cell. The Van Hove singularity near zero bias (shown in Fig. 2(d)) is not resolved because of the lower energy resolution in this large bias range. The spectra are shifted vertically for clarity.

## Note 6: STS, spectral weight transfer, and surface doping level

The STS characterization is conducted *in situ* with an Omicron low temperature STM using a tungsten tip. The density of states of the STM tip was verified to be featureless by recording a spectrum on an Au film before the actual data acquisition. dI/dV signals were obtained by superposing a 0.6 to 15 mV ac ripple (831 Hz) onto the dc tunneling voltage and detecting the first harmonic signal with a lock-in amplifier. To obtain a quantitative estimate of the doping level of each Sn√3 surface, we determined the relative intensities of the quasi-particle peak and the two Hubbard bands (UHB/LHB) by fitting the corresponding features in the $(dI/dV)/(I/V)$ curves with a series of Gaussians, as shown in Fig. S7(a)-(g). Here, we normalized the $dI/dV$ curves to remove the bias dependence of the transmission coefficient in the tunneling process. This normalization produces spectroscopic data with amplitudes that more closely resemble the spectral function of the sample [13]. Each normalized spectrum in Fig. S7(a)-(g) is obtained by dividing a $dI/dV$ curve in Fig. 2(a) by the corresponding smoothed $I/V$ data that were recorded simultaneously [14]. Finally, these spectra are normalized to the intensity of the bulk feature at ~ +1.6 V (like in Fig. 2(b)).

The broad spectral features at positive and negative sample bias shown in Fig. S4(a) correspond to the upper and lower Hubbard band, respectively. For the hole doped systems, a QPP gradually develops at ~0.1 V above the Fermi level, in between the Hubbard bands. To extract a quantitative measure of their spectral intensities, we fitted the spectra with one Gaussian representing the QPP, and two Gaussians each for the LHB and the UHB. These Gaussians are all indicated in red. Unlike the UHB, which appears to be well separated from the bulk spectral features above +1.0 V, the spectral features of the LHB overlap with those of the bulk valence band. The filled-state spectra typically show a weak minimum in the range of -0.6 to -0.8 V, indicating that the LHB extends down to this energy range. To account for the increased intensity at higher binding energy, an extra Gaussian ("VBM") is used to fit the spectra in this energy range. This Gaussian is indicated in blue and only its tail is visible in the chosen bias range. To reduce the number of free parameters in the fit, we fixed the center and width of this VBM peak at -0.7 V and 0.2 V, respectively, resulting in a VBM near -0.5 V, in good agreement with the XPS result. So altogether, we use a total of six Gaussians to fit the $(dI/dV)/(I/V)$ spectra of the hole doped surfaces: two for each Hubbard band, one for the QPP, and one for the bulk VBM. For the undoped n-0.002 sample, the Gaussian peak for the QPP is not included in the fitting. These peaks were fitted to the spectra within a limited range of -0.7 V to +0.7 V as this range contains the spectral features of √3-Sn adatom lattice. Fig. S4(a)-(g) show that reasonable fits to the $(dI/dV)/(I/V)$ data are obtained within these boundary conditions, allowing for a reliable extraction of the spectral weight of the LHB, UHB, and QPP, which is the sole purpose of this fitting exercise. Note that much of the spectral weight associated with the QPP appears above the Fermi level, indicating that all surfaces are hole-doped.

The spectral weights of the LHB, QPP and UHB, as quantified by the integrated area of the associated Gaussians, are normalized to the total spectral intensity (i.e., LHB+QPP+UHB) and plotted in Fig. 2(b). For the n-0.03, n-3, and p-0.004 samples, the normalized spectral weights originate from a fit to a single $(dI/dV)/(I/V)$ spectrum. For other substrates we recorded multiple spectra using different STM tips and averaged the normalized spectral weights extracted from those spectra. For instance, the spectral weights for the p-0.03 and B-√3 samples were averaged over eight and nine different data sets, respectively. The corresponding error bars in Fig. 2(c) indicate the standard deviation of these averages evaluated for the B-√3 sample, and applied to samples with all doping levels.

As seen in Fig. 2(b), the (normalized) spectral weights of the UHB and LHB are very similar. However, the relative spectral weight of the UHB decreases with increased hole doping of the substrate (moving to the right in the graph), while the relative spectral weight of the LHB remains constant within the margin of error. At the same time, the spectral weight of the QPP increases monotonically from 0 to ~ 0.2 with increased hole doping. In the local limit ($t = 0$), the normalized spectral weight $S$ of the QPP should vary with the doping fraction $x$ as $S = \frac{2x}{(1-x)+(1-x)+2x} = x$ [15]. The doping level derived from the spectral weight transfer could easily be overestimated by a factor of two when hopping is taken into account [15]. Indeed, our DCA calculations of the spectral weight transfer indicate that $S$ = 0.1 for x = 0.05 hole doping (calculated with the curves in Figs. 2(c)). The normalized QPP intensity in the $(dI/dV)/(I/V)$ spectrum reaches 0.2 for the B√3 substrate. We thus estimate that the maximum hole concentration in our experiments is about 0.1 hole per Sn adatom. Because the STS tunneling probability is not uniform across the entire Brillouin zone, the absolute numbers should not be taken too seriously.

We finally note that the DCA results presented in Fig. 2(c) suggest the presence of multiple components to both the LHB and UHB, supporting our use of two Gaussians for each Hubbard band. These internal UHB features, e.g. the small bump at ~0.1 eV on the "n = 1" curve in Fig. 2(c), should not be confused with the QPP feature observed experimentally in Fig. 2(a), as the "n = 1" curve in Fig. 2(c) describes the undoped Mott insulating phase. In addition, the experimental observation that most of the QPP intensity is 'stolen' from the UHB is also consistent with our DCA results. The qualitative trend of increased QPP spectral weight with stronger hole doping, as reflected by the fitting results and DCA calculations, is also clearly evident from a visual inspection of the raw data in Fig. 2(a).

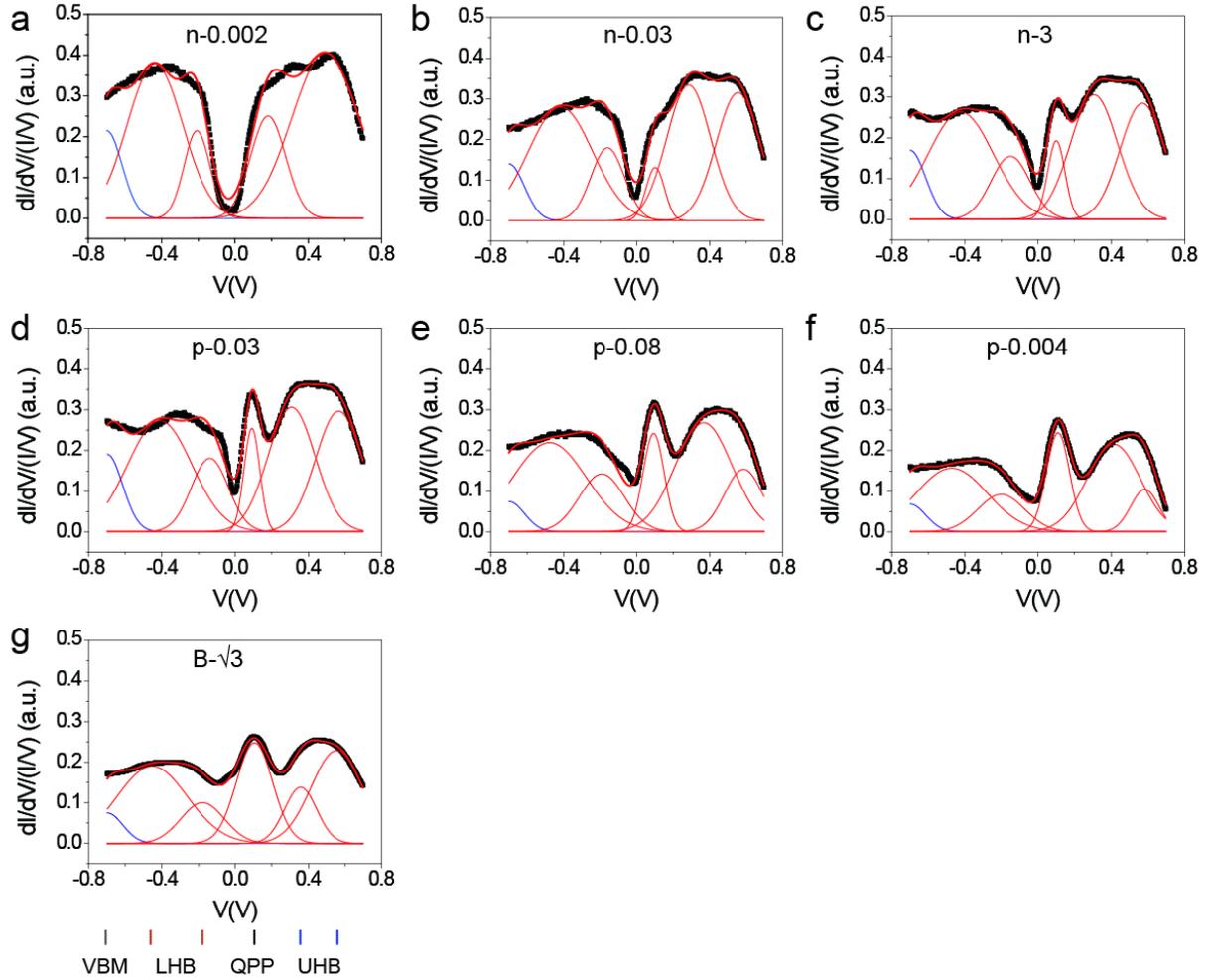

Figure S7 - Determination of the relative spectral weight of the QPP, UHB and LHB. (a-g) Fitting of the $(dI/dV)/(I/V)$ spectra using six (or five for the n-0.002 sample) Gaussian peaks. The solid bars below (g) mark the centers of the Gaussian peaks: one peak for the bulk VBM (indicated in blue), two peaks for the LHB, one peak for the QPP, and two peaks for the UHB (all indicated in red).

## Note 7: Processing procedure of the QPI data

The tunneling conductance maps $g(\mathbf{r},V) = dI/dV(\mathbf{r},V)$ were recorded from a √3-Sn surface area at 5 K, using a lock-in amplifier, at dc biases (831 Hz) of ±15 mV, ±10 mV, ±5 mV, and ±2 mV. The first row of images in Fig. S8 shows conductance maps (from the same surface area shown in Fig. 3) recorded at three different biases. The corresponding Fourier transforms, $g(\mathbf{q},V)$, are shown in the second row, after we corrected the patterns for image distortion due to the STM scanning drift, and rotated them such that a $\overline{\Gamma K}$ ($\overline{\Gamma M}$) direction runs horizontally (vertically). These $g(\mathbf{q},V)$ patterns are furthermore symmetrized according to the symmetry of the Sn√3 lattice: 3-fold rotational symmetry and mirror symmetry along the $\overline{\Gamma M \Gamma}$ axis (labelled as $q_2$ in the bottom-right panel), as shown in the third and fourth row of images in Fig.

S8, respectively.

Finally, the center of the $g(\mathbf{q}, V)$ patterns ($|\mathbf{q}| = 0$) features a larger intensity due to slow spatial variations in the $g(\mathbf{r}, V)$ maps. This enhanced intensity influences the relative intensity of the pattern as a function of distance from the center, but it is unrelated to the electronic structure. Therefore, we suppress the intensity of the middle using a Gaussian suppression [16]: the intensity is suppressed by $g'(\mathbf{q}, V) = g(\mathbf{q}, V) \times [1 - \text{Gaussian}(q = 0, \sigma)]$ with $\sigma = 1.97\pi/a$.

The resulting $g(\mathbf{q}, V)$ patterns all reveal six sharp spots. These are the Bragg spots of the √3-Sn lattice, and their presence allows us to calibrate the $\mathbf{q}$-axis. The -2 mV and 15 mV images furthermore reveal six triangular features that are located inside the lattice spots (similar feature is labelled in Fig. 3(f)). The -15 mV map is a little less resolved. Here, the outermost edges of the triangles are nearly parallel to the hexagon of Bragg spots. The -2 mV $g(\mathbf{q}, V)$ map demonstrates that these stripes are weakly split into warped segments of opposite curvature. We have extracted line profiles from these processed $g(\mathbf{q}, V)$ maps along the $\overline{\Gamma K}$ and $\overline{\Gamma M}$ directions and tracked the peak positions as shown in Fig. S9. This allows us to extract the energy dispersion of scattering vector, as shown in Figs. 3(k)&3(l). Note that all peak positions disperse continuously in the measured energy range, except for the one near the $\overline{M}$ point (marked with blue bars) in Fig. S9(b). The peak near the $\overline{M}$-point is visible only at negative bias and its dispersion shows a weak kink at about − 5 meV. This small kink is reproduced in the dispersion read from simulated QPI images (Fig. 3(l)), where the near $\overline{M}$ point peak becomes dispersionless above -4 mV. This change of dispersive behavior at about − 5 meV is related to the saddle point in the band structure near the $\overline{M}$ point, consistent with the location of the Van Hove singularity in the density of states at about − 7 meV observed with STS (Fig. 2(d)).

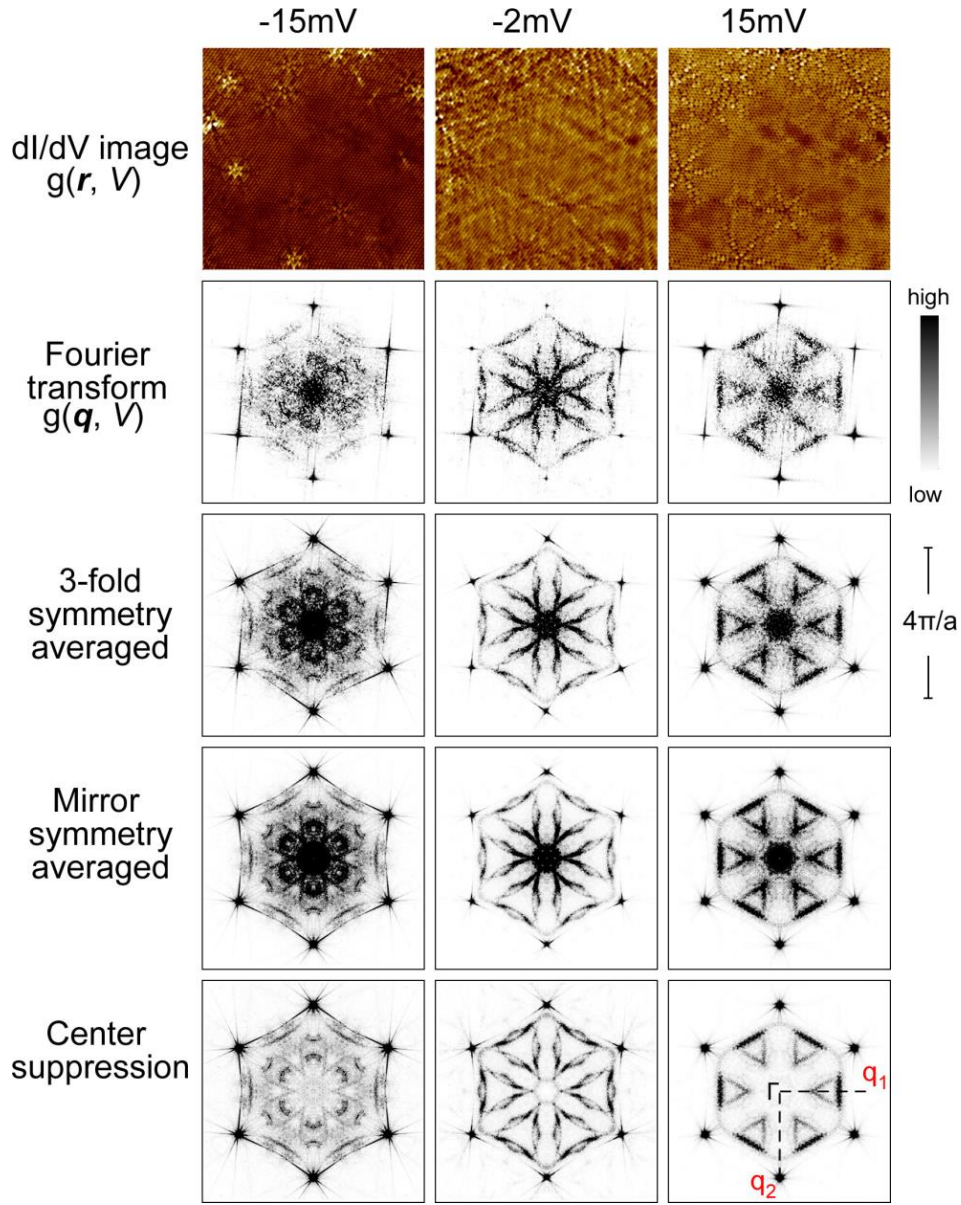

Figure S8 - Processing of the QPI data. Three step-by-step examples of the image processing applied to the QPI data. The first row shows 40 nm x 40 nm conductance maps $g(r,V)$ recorded from the same √3-Sn surface area on a B√3 substrate at 4.4 K. The corresponding STM image is shown in Fig. 3(a) of the main paper. The conductance maps contain 320 x 320 pixels, i.e. 1.25 Å/pixel. The $dI/dV$ images are taken with a lock-in amplifier using a modulation voltage of $V_{mod}$ = 1 mV for $|V_s| \geq$ 5 mV, and 0.6 mV for $|V_s|$ < 5 mV. The tunneling parameters are $V_s$ = -15 mV, $I_t$ = 1 nA; $V_s$ = -2 mV, $I_t$ = 0.2 nA; $V_s$ = 15 mV, $I_t$ = 1 nA respectively. The second row shows the power spectral density of the Fourier transforms $g(q,V)$ of the $g(r,V)$ maps. The following rows illustrate the effect of the different data processing steps on the $g(q,V)$ maps. In each $g(q,V)$ map, the color scale is adjusted in order to provide good visibility of the QPI signal within the Bragg hexagon. (The Bragg points are usually one or two orders of magnitude more intense than the maximum intensity shown in each plot). The quasi particle dispersion was measured by tracking the bright $g(q,V)$ features as a function of bias along the horizontal ($\overline{\Gamma KM}$), and vertical ($\overline{\Gamma M\Gamma}$) directions, as indicated in the bottom right QPI pattern.

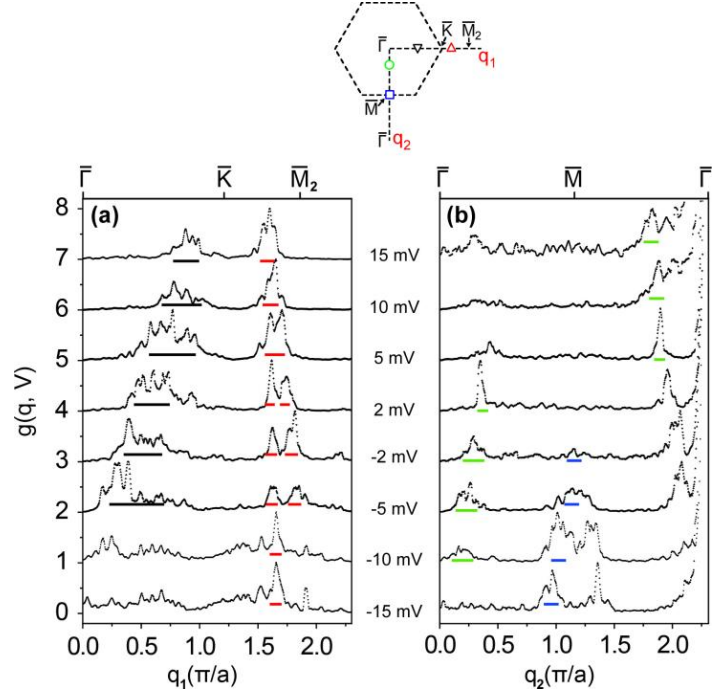

Figure S9 - Cuts of $g(\mathbf{q}, V)$ along the high symmetry directions of the surface Brillouin Zone. The surface Brillouin zone is sketched at the top, showing the direction of the cuts in the Fourier transform patterns $g(\mathbf{q}, V)$. The origin represents the center of the QPI pattern at $\mathbf{q}$ = (0, 0). Each curve is normalized to its highest intensity (the Bragg intensity near $\mathbf{q}$ = (0, 2.3π/a) in the $\overline{\Gamma M \Gamma}$ plot is excluded from this normalization). Curves are shifted vertically for clarity. The peak positions are obtained from these line profiles, as indicated by the bar below each curve, where the bar length represents the estimated error in the determination of the peak center. The rough position for those peaks are marked with hollow shapes in the sketch of the surface Brillouin zone. In the $\overline{\Gamma M \Gamma}$ plot, only one peak feature in either the Γ-M (first Brillouin zone) or M-Γ (second Brillouin zone) regions is indicated, depending on which of the two equivalent peaks is better defined. The q-vector obtained in the $\overline{M \Gamma}$ region is folded back to the $\overline{\Gamma M}$ region by q' = 4π/(√3a) – q.

## Note 8: Theoretical calculations

We performed calculations for the single-band Hubbard model on a triangular lattice using the dynamical cluster approximation (DCA) [17]. The non-interacting band-structure was taken from Ref. 18, which was obtained by downfolding of the density functional theory band structure onto an effective single-band model. Throughout, we keep the nearest neighbor hopping as $t_1 = -52.7$ meV. The on-site Hubbard interaction was then set to $U = 12.524|t_1|$ throughout [18].

The quasi-particle interference calculations were carried out using the standard T-matrix formalism [19]. Here we assume the presence of a single scatter located at the origin with a weak delta-function potential $V_0 = |t_1|/5$. The precise value of $V_0$ is not critical for our results, provide it remains weak, consistent with the homogenous nature of the measured STS spectra. The QPI intensity is given by the power spectrum of the Fourier transformed density modulations in real space $\delta\rho(\boldsymbol{q},\omega) = \frac{1}{N}\sum_{\boldsymbol{k}} \mathrm{Im} G(\boldsymbol{k}, \boldsymbol{k}+\boldsymbol{q}, \omega)$, where

$$\delta G(\boldsymbol{k},\boldsymbol{p},\omega) = G_0(\boldsymbol{k},\omega) T(\boldsymbol{k},\boldsymbol{p},\omega) G_0(\boldsymbol{p},\omega),$$

is the Green's function for the inhomogenous system, neglecting the homogeneous part; $G_0(\boldsymbol{k},\omega)$ is the noninteracting Green's function for the homogeneous system; and $T(\boldsymbol{k},\boldsymbol{p},\omega)$ is the usual T-matrix. For a single point-scatter, the T-matrix is momentum independent and given by $T(\omega) = V_0[1 - V_0 g(\omega)]^{-1}$, where $g(\omega) = \frac{1}{N}\sum_{\boldsymbol{k}} G_0(\boldsymbol{k},\omega)$.

**References:**

[1] S.M. Sze, Physics of Semiconductor Devices Ed. 3 (John Wiley & Sons, 1981).

[2] F. J. Himpsel, G. Hollinger, and R. A. Pollak, Determination of the Fermi-level pinning position at Si (111) surfaces. *Phys. Rev. B* **28**, 7014 (1983).

[3] W. Mönch, Semiconductor surfaces and interfaces Vol. 26 (Springer Science & Business Media, 2013).

[4] M. Liehr, M. Renier, R. A. Wachnik, and G. S. Scilla, Dopant redistribution at Si surfaces during vacuum anneal. *J. Appl. Phys.* **61**, 4619-4625 (1987).

[5] S. Bensalah, J-P. Lacharme, and C. A. Sébenne, Phys. Rev. B **43** 14441 (1991).

[6] J. D. Mottram, A. Thanailakis, and D. C. Northrop, J. Phys. D **8**, 1316 (1975).

[7] H. Zheng, Hao, J. Kröger, and R. Berndt, Phys. Rev. Lett. **108**, 076801 (2012).

[8] V. W. Brar et al. Nat. Phys. **7**, 43 (2010).

[9] F. Marczinowski, J. Wiebe, F. Meier, K. Hashimoto, and R. Wiesendanger, Phy. Rev. B **77**, 115318 (2008).

[10] C. Törnevik, M. Hammar, N. G. Nilsson, and S. A. Flodström, Epitaxial growth of Sn on Si (111): A direct atomic-structure determination of the (2√3×2√3)R30° reconstructed surface. *Phys. Rev. B* **44**, 13144-13147 (1991).

[11] S. Modesti, L. Petaccia, G. Ceballos, I. Vobornik, G. Panaccione, G. Rossi, L. Ottaviano, R. Larciprete, S. Lizzit, and A. Goldoni, Insulating Ground State of Sn/Si (111)–(3× 3) R 30°. *Phy. Rev. Lett.* **98**, 126401 (2007).

[12] H. Alloul, J. Bobroff, M. Gabay, and P. J. Hirschfeld, Defects in correlated metals and superconductors. *Rev. Mod. Phys.* **81**, 45 (2009).

[13] R. M. Feenstra, J. A. Stroscio, and A. P. Fein, Tunneling spectroscopy of the Si (111) 2× 1 surface. *Surf. Sci.* **181**, 295-306 (1987).

[14] P. Mårtensson, R. M. Feenstra, Geometric and electronic structure of antimony on the GaAs (110) surface studied by scanning tunneling microscopy. *Phys. Rev. B* **39**, 7744-7753 (1989).

[15] M. B. J. Meinders, H. Eskes and G. A. Sawatzky, Spectral-weight transfer: Breakdown of low-energy-scale sum rules in correlated systems. *Phys. Rev. B* **48**, 3916 (1993).

[16] M. P. Allan, A. W. Rost, A. P. Mackenzie, Yang Xie, J. C. Davis, K. Kihou, C. H. Lee, A. Iyo, H. Eisaki, and T.-M. Chuang, Anisotropic energy gaps of iron-based superconductivity from intraband quasiparticle interference in LiFeAs. *Science* **336**, 563-567 (2012).

[17] T. Maier, M. Jarrell, T. Pruschke, and M. H. Hettler, Quantum cluster theories. *Rev. Mod. Phys.* **77**, 1027 (2005).

[18] G. Li, P. Höpfner, J. Schäfer, C. Blumenstein, S. Meyer, A. Bostwick, E. Rotenberg, R. Claessen, and W. Hanke, Magnetic order in a frustrated two-dimensional atom lattice at a semiconductor surface. *Nat.*

*Commun.* **4**, 1620 (2013).

[19] S. Grothe, S. Johnston, S. Chi, P. Dosanjh, S. A. Burke, and Y. Pennec, *Phys. Rev. Lett.* **111**, 246804 (2013).